\documentclass[twocolumn,showpacs,preprintnumbers,amsmath,amssymb,floatfix,nofootinbib]{revtex4-1}

\usepackage{graphicx}
\usepackage{dcolumn}
\usepackage{bm}
\usepackage{color}
\usepackage{url}
\usepackage{amsmath}
\usepackage{aas_macros}
\usepackage{cleveref}
\usepackage{enumerate}
\usepackage{diagbox}
\crefname{section}{Sec.}{Secs.}
\crefname{equation}{Eq.}{Eqs.}
\crefname{figure}{Fig.}{Figs.}

\graphicspath{{./figures/}}
\newcommand{\be}{\begin{equation}}
\newcommand{\ee}{\end{equation}}
\newcommand{\ba}{\begin{eqnarray}}
\newcommand{\ea}{\end{eqnarray}}
\newcommand{\Mc}{{\cal M}}
\newcommand{\Ms}{M_{\odot}}

\newcommand{\T}{\vec{\Theta}}
\newcommand{\np}{\vec{\theta}}
\newcommand{\nl}{\vec{\lambda}}
\newcommand{\event}{\epsilon}

\newcommand{\cosmoP}{\vec{\Omega}}
\newcommand{\di}{\mathrm{d}}
\newcommand{\info}{\mathrm{I}}

\def\ltsima{$\; \buildrel < \over \sim \;$}
\def\simlt{\lower.5ex\hbox{\ltsima}}
\def\gtsima{$\; \buildrel > \over \sim \;$}
\def\simgt{\lower.5ex\hbox{\gtsima}}

\begin{document}

\title{Cosmological inference using only gravitational wave observations of binary neutron stars}

\author{Walter Del Pozzo}
\email{walter.delpozzo@unipi.it}
\affiliation{School of Physics and Astronomy, University of Birmingham, Edgbaston, Birmingham B15 2TT, United Kingdom}
\affiliation{Dipartimento di Fisica ``Enrico Fermi'', Universit\`a di Pisa, Pisa I-56127, Italy }
\author{Tjonnie G.F. Li}
\affiliation{LIGO - California Institute of Technology, Pasadena, CA 91125, USA }
\affiliation{Department of Physics, The Chinese University of Hong Kong, Shatin, N.T., Hong Kong}
\author{Chris Messenger}
\affiliation{SUPA, School of Physics and Astronomy, University of Glasgow, Glasgow G12 8QQ, United Kingdom}
\date{\today}

\begin{abstract}
Gravitational waves emitted during the coalescence of binary neutron
star systems are self-calibrating signals. As such, they can provide a
direct measurement of the luminosity distance to a source without the
need for a cross-calibrated cosmic distance-scale ladder. In general, however, the
corresponding redshift measurement needs to be obtained via
electromagnetic observations since it is totally degenerate with the total mass
of the system. Nevertheless, Fisher matrix studies
have shown that, if information about the equation of state of the
neutron stars is available, it is possible to extract redshift
information from the gravitational wave signal alone. Therefore,
measuring the cosmological parameters in pure gravitational-wave
fashion is possible. Furthermore, the huge number of sources potentially
observable by the Einstein Telescope has led to speculations that the
gravitational wave measurement is potentially competitive with
traditional methods. The Einstein Telescope is a conceptual study for
a third generation gravitational
wave detector which is designed to yield $10^3-10^7$ detections of
binary neutron star systems per year. This study
presents the first Bayesian investigation of the accuracy with which
the cosmological parameters can be measured using information coming \emph{only} from the 
gravitational wave observations of binary neutron star systems by Einstein Telescope. 
We find, by direct simulation of $10^3$
detections of binary neutron stars, that, within our simplifying
assumptions, $H_0,\Omega_m,\Omega_\Lambda,w_0$ and $w_1$ can be
measured at the $95\%$ level with an accuracy of $\sim
8\%,65\%,39\%,80\%$ and $90\%$, respectively. We also find, by
extrapolation, that a measurement accuracy comparable with current
measurements by Planck is possible if the number of gravitational 
wave events observed is $O(10^{6-7})$.We conclude that, while not competitive with 
electro-magnetic missions in terms of significant digits, gravitational wave alone are 
capable of providing a complementary determination of the dynamics of the Universe.
\end{abstract}

\maketitle

\section{Introduction} 
The family of second generation interferometers Advanced LIGO~\cite{urlAdvLIGO}
began its operations in the last quarter of 2015 \cite{GW150914-DETECTORS}.
Advanced Virgo~\cite{urlAdvVirgo} is scheduled to join the LIGO network in 
2017, with KAGRA ~\cite{Kuroda2010a} and LIGO India~\cite{LIGOIndia} to follow
afterwards. The detection of gravitational waves from the coalescence of 
merging black holes \cite{GW150914-DETECTION,GW151226-DET,BBH-O1}
led already to important scientific measurements as tests of general relativity
~\cite{GW150914-TESTOFGR,BBH-O1} and astrophysics \cite{GW150914-PARAMESTIM,GW150914-ASTRO,BBH-O1}.
Given the expected number of yearly
detections~\cite{GW150914-RATES,BBH-O1,Abadie2010a}, the expectations on the scientific deliverables
are high: tests of the strong field dynamics of General
Relativity \cite{GW150914-TESTOFGR,BBH-O1,DelPozzo2011a,Li2012a,Cornish2011a}; a ``cosmic distance
scale ladder''-free determination of the Hubble
constant~\cite{Nissanke2010a,DelPozzo2012a,Taylor2012a}; a determination of the
neutron star equation equation of state
\cite{Read2009a,Hinderer2010a,Lackey2014a,DelPozzo2013a}.

Detectors beyond the second generation are already being envisaged. For
instance, the Einstein gravitational wave Telescope (ET) \cite{ETconcept} is a
proposed underground detector consisting of three 10 km arm-long Michelson interferometers in a
triangular topology with opening angles of 60 degrees \cite{Freise2009a}. The
strain sensitivity is estimated as factor 10 better than second generation
detectors, down to frequencies of $1-3$ Hz depending on the actual
configuration of the instrument 
\cite{Punturo2010b}. The high sensitivity promises the detection of a
very large number of gravitational waves (GW) signals with large
signal-to-noise ratios (SNR), thus allowing for unprecedented
population studies as well as extremely accurate measurements of the
physical parameters of coalescing binary systems \cite{ETconcept}.

\subsection{Cosmological inference with gravitational waves}
When estimating the parameters of GW sources, and in
particular the coalescences of binary neutron stars and black holes, the
luminosity distance can be observed directly
\cite{Schutz1986a,Sathyaprakash2009a}. This makes GW an ideal laboratory to
place samples in the Hubble diagram in a manner that is free from the potential systematics
affecting electromagnetic (EM) methods. Unfortunately, in the vast majority of
the cases, the redshift cannot be measured from GW alone and this piece of
information needs to be extracted by means other than GW.

In the recent years, various proposals have been put forward to overcome this
difficulty.  For instance, one can assume that the coalescences of
compact binaries are the progenitors of short gamma ray bursts (sGRB)
\cite{Nakar:2005cz}. In this case, coincident observations of
a GW event and the correspondent sGRB would allow the measurement of the luminosity
distance from GW and the redshift from spectroscopy of the host
galaxy\cite{Nissanke2010a,Sathyaprakash2010a,Zhao2011a,Nissanke2013a}\footnote{Kilonovae are also expected EM counterparts to BNS coalescences~\cite{MetzgerEtAl:2010}. However their utility as cosmological probes is yet unclear due to their intrinsically faint luminosities~\cite[e.g.][]{Tanaka:2016}  
which limits the distance at which they can be confidently detected.}.  For
second generation interferometers, this method indicates a relative accuracy on
the measurement of the Hubble constant $H_0$ of few percent in the
case where 10-15 such events are detected.  However, whether the 
coalescences of compact binaries are the
progenitors of sGRBs is still a matter of debate. Also, the fraction of GW
events also observable as sGRBs might be as low as
$10^{-3}$\cite{Belczynski2008a} due to sGRB beaming effects. 

An alternative approach, following broadly the argument first given in
Ref.~\cite{Schutz1986a}, would be to statistically identify the possible host
galaxies of a GW event to obtain a distribution of possible redshifts
associated with each GW detection.  This method should yield $\sim$ 5\% percent accuracy
on $H_0$ using 20-50 events~\cite{DelPozzo2012a} observed by Advanced
LIGO/Virgo. A similar methodology has also been applied to space-based
detectors~\cite{MacLeod2008a,Petiteau2011a}.

A few methods aim at extracting the redshift using GW observations alone. For
example, one can use the knowledge of the (rest frame) mass function
of NS and the measured (redshifted) mass to infer the redshift of the
source \cite{Taylor2012a,Taylor2012b}. In this framework, second generation
interferometers should infer the Hubble constant $H_0$ with $\sim$ 10\% accuracy
using about 100 events \cite{Taylor2012a}.

The results of advanced interferometers can be greatly improved by third
generation instruments such as ET. In fact, ET can
probe regions of the Universe where the effects of the dark energy will be
substantial, thus allowing an independent sampling of the cosmic history.

The potentialities of ET have already been investigated by various groups
\cite{Sathyaprakash2010a,Zhao2011a,Sathyaprakash:2011}, concluding that, when only a limited set
of cosmological parameters is considered, the accuracy of the inference is
comparable to that of current EM measurements. 

\subsection{Outline}
In this paper, we will expand on the approach proposed by Messenger
and Read \cite{Messenger2012a}
in which if one of the two components is a NS, information about the equation of
state (EOS) allows a direct measurement of the rest-frame masses and thus of the
source redshift \cite{Messenger2012a}. Using Fisher matrix formalism, the authors
estimate the accuracy with which $z$ can be measured to be  $\sim 8 - 40\%$,
depending on the EOS and on the distance to the source. Recently, a similar
investigation was carried out in \cite{Li2013a} using a more realistic Monte Carlo
data analysis method. The authors concluded that the average uncertainty is
closer to $40\%$ for a hard EOS and essentially independent of redshift.

Nevertheless, given the large number of sources that can be observed by ET and
the possibility of combining information across them, even the large
uncertainty reported in Ref.~\cite{Li2013a} could be sufficient to obtain
interesting indications on the accuracy with which ET will measure the
cosmological parameters. In this paper, we explore this idea in a
simplified scenario and conclude that ET can indeed set bounds
that are comparable to current EM measurement. We are interested in the 
cosmological information that can be inferred \emph{exclusively} from the observation of gravitational waves. 
We will thus not discuss the potential of coincident GW-EM detections 
which are presented elsewhere \cite{Zhao2011a,Sathyaprakash:2011}. We note here that, 
because of the co-location of the three ET interferometers and because of its topology, 
its expected sky resolution is extremely poor.
Consequently, the probability of a successful EM-GW association is a priori very small. 
Note that at the time of ET, second generation detectors are expected to be operational with improved sensitivities~\cite{threeG}.
For a substantial fraction of the loudest GW events, the sky localisation from a network made of ET and advanced detectors 
will be vastly improved compared to ET alone. In this case, some of the aforementioned EM+GW methods might become feasible and used to yield 
constraints on the cosmological parameters. 

The rest of the paper is organised as follows. In \cref{s:method} we cast the
problem in a Bayesian framework, and identify the necessary components to
arrive at the cosmological inference. In \cref{s:simulation} we describe the
procedures of simulating GW events and the detector noise, and the
implementation of the analysis. In \cref{s:results} we present the results of
our simulations and finally in \cref{s:discussion} we summarise and discuss our
results. The mathematical solution to the problem of the inference of
the cosmological parameters in the presence of a detection threshold
is given for completeness in Appendix \ref{app:A}.

\section{Method}\label{s:method} 

In this section we present a Bayesian solution to the problem of computing
posterior probability density functions for a set of cosmological parameters
from GW data. We broadly follow the presentation in
\cite{DelPozzo2012a}. Note that the treatment is not specific to the
case of ET.

\subsection{Inference of the cosmological parameters in the absence of
a detection threshold}

Consider a catalogue of GW events
$\mathcal{E}\equiv\{\event_1,\ldots,\event_N\}$. Each event is defined as a
stretch of data $d_i(t)$ given by the sum of noise $n_i(t)$ and a gravitational
wave signal $h_i(\vec{\Theta_i};t)$, \textit{i.e.}
\begin{align}
	\event_i\, :\, d_i(t)=n_i(t)+h_i(\T_i;t)\,,
\end{align}
where $\T_i$ indicates the set of all parameters of the signal $i$. 

The noise is taken to be a stationary Gaussian process with a zero mean and covariance
defined by its one-sided spectral density $S_n(f)$ such that 
\begin{align}
	p(n_i|\info) & \propto\exp\left\{-\frac{1}{2}\int_0^{\infty}\di f\, 4 \frac{|\tilde{n_i}(f)|^2}{S_n(f)} \right\} \,,\nonumber \\
	& \propto \exp\left\{ -\frac{1}{2} \left( n\,\right| \left. n \right) \right\}
\end{align}
where $\info$ represents all the relevant information for the inference
problem, a tilde represents the Fourier transform, and where we have introduced
a scalar product between two real functions $A(t)$ and $B(t)$ as
\begin{align}
	\left( A\right|\left. B \right) = 4 \Re \int^{\infty}_{0} \di f\, \frac{\tilde{A}^*(f) \tilde{B}(f)}{S_n(f)}.
	\label{eq:scalarproduct}
\end{align}
The likelihood of observing the event $\event_i$ is then given by
	\begin{align} \label{eq:likelihood}
		p(\event_i|\T_i,\mathrm{S},\info) & \propto \exp\left\{ -\frac{1}{2} \left( d_i-h_i \,\right|\left. d_i-h_i \right) \right\}
	\end{align}
where $\mathrm{S}$ is the signal model that relates the signal parameters
$\vec{\Theta}_i$ to a gravitational wave signal $h$. Moreover, the
signal-to-noise ratio (SNR) $\rho$ can be succinctly written as
\begin{align}\label{eq:snr}
	\rho=\sqrt{ \left( h\,\right| \left. h \right)}\,.
\end{align}
The posterior distribution for any parameter in our signal model $\mathrm{S}$
is related to the likelihood in \cref{eq:likelihood} through the application of
Bayes' theorem
\begin{align} \label{eq:posterior}
	p(\T_i|\event_i,\mathrm{S},\info) \propto p(\T_i|\mathrm{S},\info) p(\epsilon_i | \T_i,\mathrm{S}, \info)
\end{align}
where $p(\T_i|\mathrm{S},\info)$ is the prior probability distribution for the
parameters $\T_i$.  When multiple independent detectors are included in the
analysis, the likelihood (\cref{eq:likelihood}) generalises to
\begin{align}
	p(\event_i|\T_i,\mathrm{S},\info)=\prod_k p(\event_i^{(k)}|\T_i,\mathrm{S},\info)\,.
\end{align}
For this work, we are only interested in the posterior probability for a subset
of parameters $\cosmoP\equiv
\left\{H_0,\Omega_m,\Omega_\Lambda,\ldots\right\}$. Therefore, we marginalise
over the remaining subset of parameters $\np_i$, \textit{i.e.}
\begin{align}
	p(\cosmoP|\event_i,\mathrm{S},\info)&=\int \di\np_i\, p(\T_i|\event_i,\mathrm{S},\info)\nonumber\\
	&=\int \di\np_i\, p(\np_i,\cosmoP|\mathrm{S},\info)p(\event_i|\np_i,\cosmoP,\mathrm{S},\info)\nonumber\\
	&=p(\cosmoP|\mathrm{S},\info)\int \di\np_i\, p(\np_i|\cosmoP,\mathrm{S},\info)p(\event_i|\np_i,\cosmoP,\mathrm{S},\info) \nonumber \\
	&=p(\cosmoP|\mathrm{S},\info) \mathcal{L}(\event_i,\cosmoP) \,,
\end{align} 
Where we have introduced the so-called ``quasi-likelihood'' \cite{Jaynes2003a}
\begin{align} \label{eq:quasilikelihood}
	\mathcal{L}(\event_i,\cosmoP) \equiv\int \di\np_i\, p(\np_i|\cosmoP,\mathrm{S},\info)p(\event_i|\np_i,\cosmoP,\mathrm{S},\info).
\end{align}
Finally, the posterior for $\cosmoP$ given an ensemble of events $\mathcal{E}$
can be shown to be
\begin{align} \label{eq:joint-posterior}
	p(\cosmoP|\mathcal{E}, \mathrm{S},\info) =p(\cosmoP|\mathrm{S},\info) \prod_i \mathcal{L}(\event_i,\cosmoP).
\end{align}
Therefore, in order to obtain the posterior for $\cosmoP$, we need to perform a
multi-dimensional integral in \cref{eq:quasilikelihood} for each of the GW
events. The description of this procedure and the generation of data follow in
\cref{s:simulation}.

\section{Analysis}\label{s:simulation}

In this section we describe the simulation that was performed.
Firstly, we outline the generation of the data, consisting of the GW signal
model, the astrophysical and cosmological assumptions regarding the source
population, and the simulation of the detector noise. Secondly, we show the
data analysis implementation with which the simulated data was analysed. In particular, we describe the construction of the
quasi-likelihood, and it subsequent use to arrive at our cosmological
inference. The GW signals, and the detector noise have been generated using the
LIGO Analysis library (LAL)~\cite{urlLAL}.

\subsection{Astrophysical and cosmological assumptions}\label{s:AFM}

The NS masses are distributed according to a Gaussian distribution
with a mean of $1.35M_{\odot}$ and a standard deviation of $0.15M_{\odot}$
\cite{Kiziltan2010a} which is assumed constant throughout the cosmic
history.
For the NS equation of state we consider three cases; a hard EOS, a medium
and a soft EOS. They are labelled as MS1~\cite{ms1}, H4~\cite{h4} and SQM3~\cite{sqm3}. We investigate these
three cases since in~\cite{Messenger2012a} it was shown that the accuracy with
which the redshift can be measured depends on the magnitude of the physical effects related to
the details of the EOS.  One can think of these three cases as an optimistic, a
realistic and a pessimistic one, respectively.  

The events are distributed uniformly in the cosine of the inclination, polarisation and time of
arrivals.  The events are also uniformly distributed in comoving volume. Their
redshifts are sampled from the probability density given by~\cite{Coward2005a} 
\begin{align} \label{eq:pz1}
	p(z|\cosmoP) = \frac{\di R(z)}{\di z}\frac{1}{R(z_{\mathrm{max}})}
\end{align}
where $R(z)$ is the cosmic coalescence rate. It is worth nothing that
$p(z|\cosmoP)$ is an explicit function of $\cosmoP$. 
The differential cosmic coalescence rate is equal to
\begin{align} \label{eq:pz2}
	\frac{\di R(z)}{\di z} = \frac{\di V}{\di z}\frac{r_0 e(z)}{1+z}
\end{align}
where $r_0$ is the local rate, $e(z)$ is the cosmic star formation rate and $V$
is the comoving volume.  In a Friedmann-Robertson-Walker-Lemaitre
(FRWL) universe, the
rate of change of $V$ with $z$ is given by
\begin{align} \label{eq:pz3}
	\frac{\di V}{\di z} = 4\pi\frac{D_L^2(z)}{(1+z)^2 H(z)}\,,
\end{align}
where we have introduced the Hubble parameter
\begin{align}
	H(z)=H_0\sqrt{\Omega_m (1+z)^3+\Omega_k (1+z)^2+\Omega_\Lambda E(z,w(z))}
\end{align}
and the \emph{luminosity distance} \cite{Hogg1999a}
\begin{align} \label{eq:dl}
	D_L(z)=\left\{\begin{array}{ll}
		\frac{(1+z)}{\sqrt{\Omega_k}}\sinh[\sqrt{\Omega_k}\int_0^z\frac{dz^\prime}{H(z^\prime)}] & \mbox{for \, $\Omega_k > 0$} \\
		(1+z)\int_0^z\frac{dz^\prime}{H(z^\prime)} & \mbox{for \, $\Omega_k = 0$} \\
		\frac{(1+z)}{\sqrt{|\Omega_k|}}\sin[\sqrt{|\Omega_k|}\int_0^z\frac{dz^\prime}{H(z^\prime)}] & \mbox{for \, $\Omega_k < 0$} \end{array} \right.
\end{align}
$H_0$ is the Hubble constant, $\Omega_m$ is the matter fractional density,
$\Omega_\Lambda$ is the fractional energy density of dark energy,
$\Omega_k=1-\Omega_m-\Omega_\Lambda$ is the curvature.
Finally 
\begin{align}\label{eq:DE}
	E(z,w(z))=(1+z)^{3(1+w_0+w_1)} e^{-3w_1z/(1+z)}
\end{align}
is a convenient parametrisation to capture the effects of the redshift
evolution of dark energy \cite{Linder2003a}.  For $\cosmoP$ we chose fiducial
values of
\begin{align}\label{eq:fiducial-values}
	(h,\Omega_m,\Omega_\Lambda,\Omega_k,w_0,w_1)^{\rm fid}=(0.7,0.3,0.7,0,-1,0),
\end{align}
where $h=H_0/100$km$\cdot$Mpc$^{-1}\cdot$s$^{-1}$. Even though ET
horizon distance is $\simeq 37$ Gpc ($z\simeq 4.15$ for our fiducial cosmology), we limit our analysis to
$z_{\mathrm{max}}=2$ as this corresponds approximately the sky averaged horizon
distance of $13$ Gpc for BNS systems \cite{ETMDC}.
For simplicity, we decided to assume a star formation rate $e(z)$ that
does not change with redshift and is therefore irrelevant for our problem.

\begin{figure}[h]
	\includegraphics[width=\columnwidth]{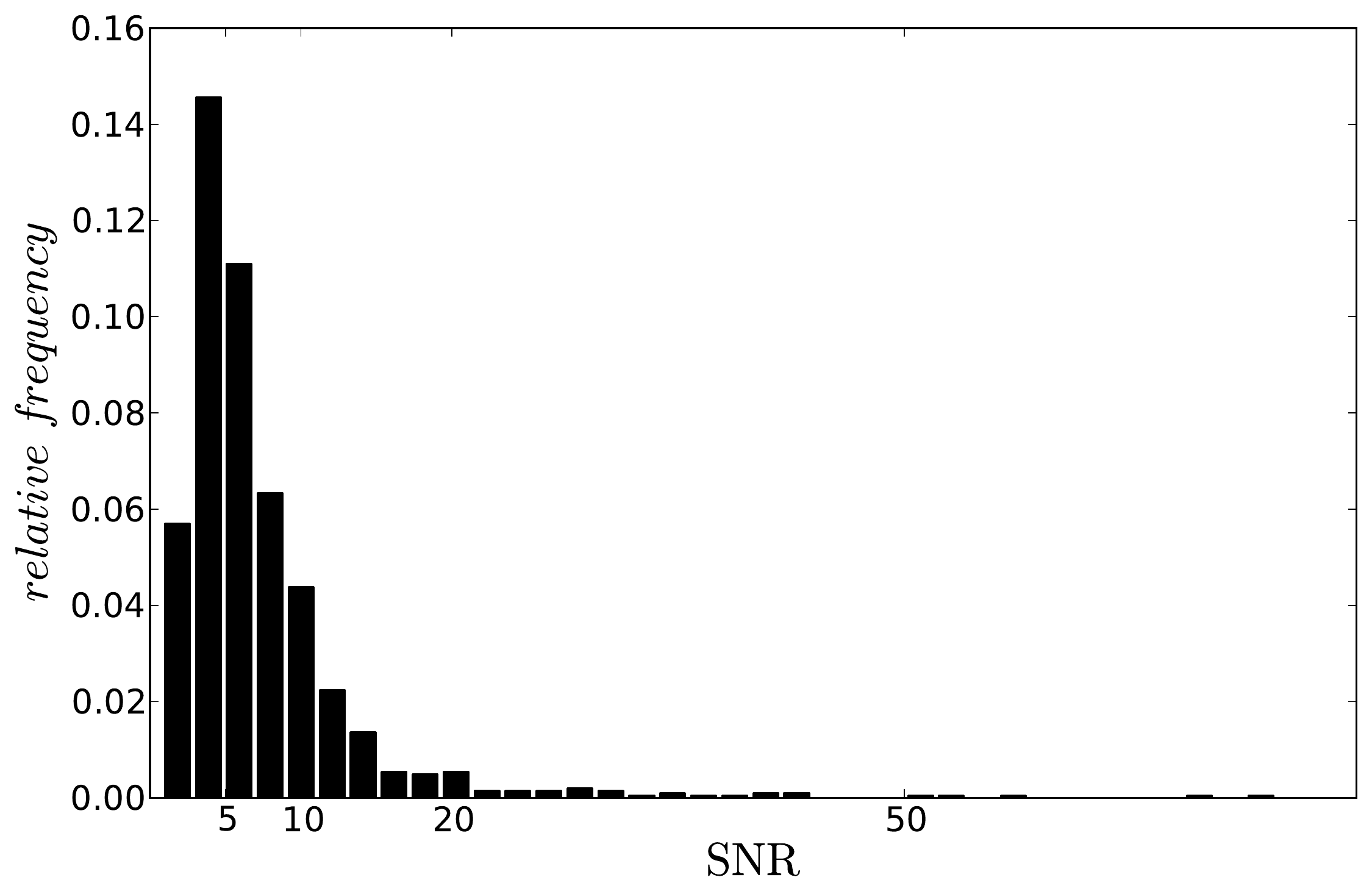}
	\caption{Network SNR distribution of the 1,000 BNS events
		generated sampling \cref{eq:pz1,eq:pz3} for a our
                fiducial cosmology \cref{eq:fiducial-values}.}
	\label{fig:inj-distribution}
\end{figure}

We simulated 1,000 binary NS events as observed by ET. The parameters $\theta_i$ of each
individual source have been generated according to the assumptions described in
\cref{s:AFM}. The corresponding waveform $\tilde{h}(f,\T_i)$ was then
added to Gaussian noise which is coloured according to the amplitude spectral density shown in
\cref{fig:etb}.  In \cref{fig:inj-distribution} we show the network SNR
distribution in the ET detector computed using \cref{eq:snr}. 
%
%

Differently from most existing literature, we do not filter our sources with
any SNR threshold. If we were to do so, we would be introducing a selection
bias \cite{MessengerVeitch:2013}. 
Note that, due to the potentially large number of sources observed by
ET and their distribution in co-moving volume, the vast majority of
them will in practice not be detected in a search which uses the
SNR as decision statistics. It is known that
ignoring these unregistered sources leads to a significant bias in the
estimation of ``global'' parameters, see \cite{MessengerVeitch:2013}.

The main reason for the emergence of biases is intimately linked
to the functional form for the prior on $z$
\cref{eq:pz1}. Since \cref{eq:pz1} quantifies the prior expectation
regarding the distribution of sources in the co-moving volume, it is
an explicit function of $\cosmoP$. 
The quasi-likelihoods for the majority of our simulated events are almost uniform in $\cosmoP$, see Section
\ref{s:cosmo-inference}, therefore our inference is greatly influenced by the
prior distribution: if one were to analyse sources that are louder
than some threshold SNR the overall population of events would appear
on average closer than the actual population. At the same time, the
observed distribution of $z$ would follow the actual cosmological
distribution. Since $p(z|\cosmoP)$ \cref{eq:pz1} relates $D_L$ and
$z$ via $\cosmoP$, this leads for estimates of $\Omega_m
\rightarrow 1$ and $h \rightarrow 0$. Similarly, if one were to
consider only events that are quieter than a given SNR threshold, $\Omega_m
\rightarrow 0$ and $h \rightarrow 1$, thus $\Omega_\Lambda
\rightarrow 1$. 
In Appendix \ref{app:A}, we give
a mathematical solution to the problem of inferring $\cosmoP$ that
accounts for sets of unobserved events. However, the solution in
\ref{app:A} is not computationally treatable with current techniques
and for a large number of events, therefore
we opted for an SNR selection threshold of 0 and analysed all
simulated events. 
Furthermore, our choice relies on the capacity of distinguishing
between low SNR GW signals and low SNR background events due to noise
in the detector. A discussion of this problem can be found in
\cite{MessengerVeitch:2013}. 
The triangular configuration of ET provides an additional tool to
study the distribution of signal and noise low SNR events. Thanks to
its topology, ET admits the construction of a null stream which is
devoid of any GW signal as the sum of the outputs of the individual Michelson
detectors \cite{Freise2009a}. Being a pure noise process, the analysis
of the null stream can be used to understand the SNR distribution of
noise events which can then be used to infer the SNR distribution of
quiet sources.

\subsection{The signal model} \label{s:signalmodel}
In the previous paragraph we introduced the signal model $\mathrm{S}$ without
specifying its properties. In this section, we lay out the assumptions that go in
the construction of $\mathrm{S}$.

In modelling the GW from a binary system, we limit our analysis to
the inspiral phase of the coalescence process. We model the inspiral using an
analytical frequency domain 3.5 post-Newtonian waveform in which we ignore
amplitude corrections and the effects of spins. This is not a big limitation as
NS are expected to be slow rotators\cite{OShaughnessy2008a}.  In
particular, we use the so-called \texttt{TaylorF2} approximant
\cite{Buonanno2009a}, which can be written as
\begin{align}
	\tilde{h}(\T;f)=A(\T)f^{-7/6}e^{i\Phi(\T;f)}\,,
\end{align}
where the waveform is written in terms of the amplitude $A(\T)$ and the phase
$\Phi(\T;f)$.

The amplitude of the waveform $A(\T)$ is given by
\begin{align}
	A(\T)\propto \frac{\Mc^{5/6}}{D_L}Q(\iota,\psi,\alpha,\delta)\,
\end{align}
where we have introduced the chirp mass $\Mc =
m_1^{3/5}m_2^{3/5}/(m_1+m_2)^{1/5}$, $D_L$ is the luminosity distance defined
in \cref{eq:dl}, $\left( \alpha, \delta \right)$ signify the sky position of
the source, and $\left( \iota, \psi \right)$ give the orientation of the binary
with respect to the line of sight \cite{Buonanno2009a}.

The wave phase can be written in the form
\begin{align} \label{eq:phasePP}
	\Phi(\T;f) =\; & 2\pi f t_c - \phi_c - \frac{\pi}{4} \nonumber \\ 
	& +\sum_{n=0}^7\left[\psi_n+\psi_n^{(l)}\ln f\right]f^{(n-5)/3},
\end{align}
where the $\psi_n$ are the so-called
post-Newtonian coefficients (see \textit{e.g.} \cite{Mishra2010a}), which are
functions of the component masses $m_1$ and $m_2$, and $(t_c,\phi_c)$ are the
time and phase of coalescence. Note that all masses are defined in the observer
frame, and the rest frame mass $m_{\mathrm{rest}}$ is related to the observed
mass through 
\begin{align}
	m = m_{\mathrm{rest}}(1+z)\,,
\end{align}
where $z$ is the redshift of the GW source.

The description of the phase in \cref{eq:phasePP} assumes that the object is a
point particle, and thus cannot be tidally deformed. However, since we consider
all of our events are binary NS coalescences, we modify the
gravitational wave phase in \cref{eq:phasePP} by including the finite-size
contributions to the phase. These in turn depend on the \emph{tidal
deformability} $\lambda(m_{\mathrm{rest}})$ \cite{Hinderer2010a} of the star which is a
function of its equation of state and its rest frame mass. The finite-size
contributions to the GW phase, as a function of observed masses, is given by
\begin{widetext}
	\begin{align} \label{Phitidal}
		\Phi_{\rm tidal}(\T;f) 
		= \sum_{a=1}^2 \frac{3\lambda_a (1+z)^5}{128 \eta M^5}\left[ -\frac{24}{\chi_a} \left(1 + \frac{11 \eta}{\chi_a}\right)\,(\pi M f)^5 -\frac{5}{28 \chi_a} \left(3179 - 919\,\chi_a - 2286\,\chi_a^2 + 260\,\chi_a^3 \right)\,(\pi M f)^7
		\right],
	\end{align}
\end{widetext}
where the sum is over the components of the binary, $\chi_a = m_a/M$,
$\lambda_a = \lambda(m_a)$ where $m_a$ are the component masses, $M$ is the
total mass, and $\eta = m_1 m_2/M^2$. 

Knowledge of the EOS and using information encoded in the GW tidal
phase contribution allows to measure the redshift of the source
\cite{Messenger2012a}.  While the EOS
is not known yet, various studies have shown that it could be possible to infer
it from observations of BNS with second generation detectors
\cite{Lackey2014a,Read2009a,Hinderer2010a,Markakis2010a,Damour2012a,DelPozzo2013a}.
In what follows, we will assume that the nature of the NS interior is
known. 

\subsection{Data analysis}
\begin{figure}[h] 
	\includegraphics[width=\columnwidth]{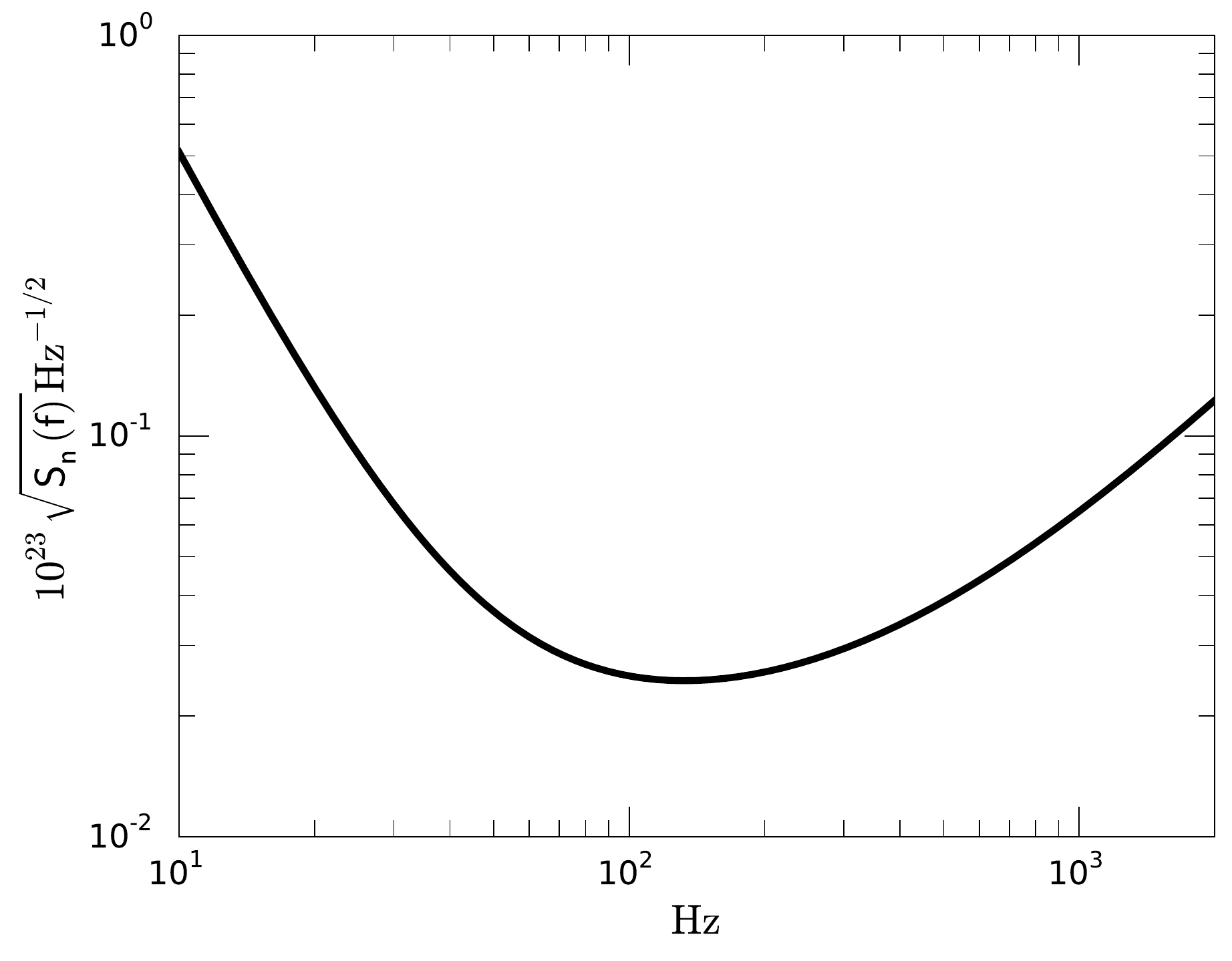}
	\caption{Amplitude spectral density for ET in the ``B'' configuration.}
	\label{fig:etb} 
\end{figure}

For our analysis, we assumed a noise curve for ET
corresponding to the ``B'' configuration \cite{Hild2008a}, corresponding to the projected
sensitivity achievable with the current technologies
\cref{fig:etb}.  Given the anticipated rates of compact binary coalescences
\cite{Abadie2010a}, the detection rates of binary NS systems in ET
are expected to lie in the range $10^3-10^7$ yr$^{-1}$\cite{ETconcept}. 
%
%

The parameters $\vec{\theta}$ of our signal model are the component
masses $m_1$ and $m_2$, inclination $\iota$, polarisation $\psi$, right
ascension $\alpha$ and declination $\delta$, the time of coalescence $t_c$, the
phase of coalescence $\phi_c$, luminosity distance $D_L$ and the redshift $z$. In
our analysis we ignored the presence of spins, as it is believed to be
small in binary NS systems\cite{OShaughnessy2008a}. We analysed our ensemble of sources assuming that the EOS of the NS is known, thus accounting to a total of three
analysis runs (one for each of our predefined hard, medium and soft
EOSs). To obtain the posterior probability distribution on the
cosmological parameters $\cosmoP$ we proceeded in three steps.
Firstly, we analysed each source to compute a quasi-likelihood as a function of
the redshift $z$ and the luminosity distance $D_L$. Secondly, these
quasi-likelihoods are then converted into quasi-likelihoods as a function of
the cosmological parameters $\cosmoP$ as shown in \cref{eq:quasilikelihood}.
Finally, the posterior probability function for the cosmological parameters
given an ensemble of events are computed from \cref{eq:joint-posterior}.

\subsubsection{Obtaining the quasi-likelihood}
For each event $\event_i$, we compute the quantity
\begin{align}\label{eq:step1-likelihood}
\mathcal{L}(\event_i,D_L,z,\cosmoP) \equiv\int \di\nl\, p(\nl|\cosmoP,\mathrm{S},\info)p(\event_i|\nl,\cosmoP,\mathrm{S},\info)
\end{align}
that is a partially marginalised quasi-likelihood, where the
marginalisation is done on all parameters that are not relevant to the
inference of $\cosmoP$. These are $\nl \equiv (m_1,m_2,\psi,\iota,\phi_c,t_c,\alpha,\delta)$.
The further marginalisation over $z$ and $D_L$ will be
described later on. For the time being, let us describe the details of the
analysis for the computation of \cref{eq:step1-likelihood}. The above
integral was computed using a Nested Sampling algorithm \cite{Skilling2004a}
implemented similarly to what described in \cite{Veitch2010a}.  For
each of the three analysis runs, we chose the same prior probability
distributions for all parameters, with the exception of the component masses.
For the common parameters we used uniform probability distributions on the
2-sphere for sky position $(\alpha,\delta)$ and orientation $(\psi,\iota)$ and
uniform in the time of coalescence $t_c$ with a width of 0.1 seconds around the
actual coalescence time. For the first marginalisation, we choose
uniform sampling distributions for both $D_L$ and $z$ in the intervals
[$1,10^5$] Mpc and [0,2], respectively.

For the component masses, the priors were
different across the different runs; each EOS in fact
predicts not only the functional form of the tidal deformability $\lambda(m)$
that enters in the phase of the GW waveform, but also the
maximum permitted mass of the NS itself. Therefore, for the three
EOSs under consideration we used maximum expected rest frame mass
$M_{\mathrm{max}}$ of $2.8,2.0,2.0\Ms$ for MS1, H4 and SQM3 respectively. The
prior probability distribution for the component masses was then chosen to be
uniform between 1$\Ms$ and $M_{\mathrm{max}}$.

\subsubsection{Cosmological inference}\label{s:cosmo-inference}
The marginalisation over the redshift and luminosity distance was then
performed as follows: once a cosmological model is introduced, $z$ and $D_L$
are not independent parameters anymore, but they are related unequivocally by
\cref{eq:dl}, thus -- after some algebra which can be found in
\cite{DelPozzo2012a} -- we are left with the following integral to compute
\begin{align} \label{eq:step2-likelihood}
	\mathcal{L}(\event_i,\cosmoP) =\int_0^{z_{\mathrm{max}}} \di z\, p(z|\cosmoP,S\info)\mathcal{L}(\event_i,D_L(z),z,\cosmoP)\,,
\end{align}
where $p(z|\cosmoP,S\info)$ is given in \cref{eq:pz1} and we chose,
consistently with the sources generation, $z_{\mathrm{max}}= 2$.  

One of the problems we needed to overcome in order to perform the integral in
\cref{eq:step2-likelihood} was how to represent $\mathcal{L}(\event_i,D_L(z),z,\cosmoP)$
in a tractable way.  In fact, one of the outputs
of the Nested Sampling algorithm is a set of samples drawn from the integrand
in \cref{eq:step1-likelihood} which is difficult to manipulate -- in particular
difficult to integrate -- without making any assumptions about the
underlying probability distribution. 

\begin{figure*}[htp!]
	\includegraphics[width=\columnwidth]{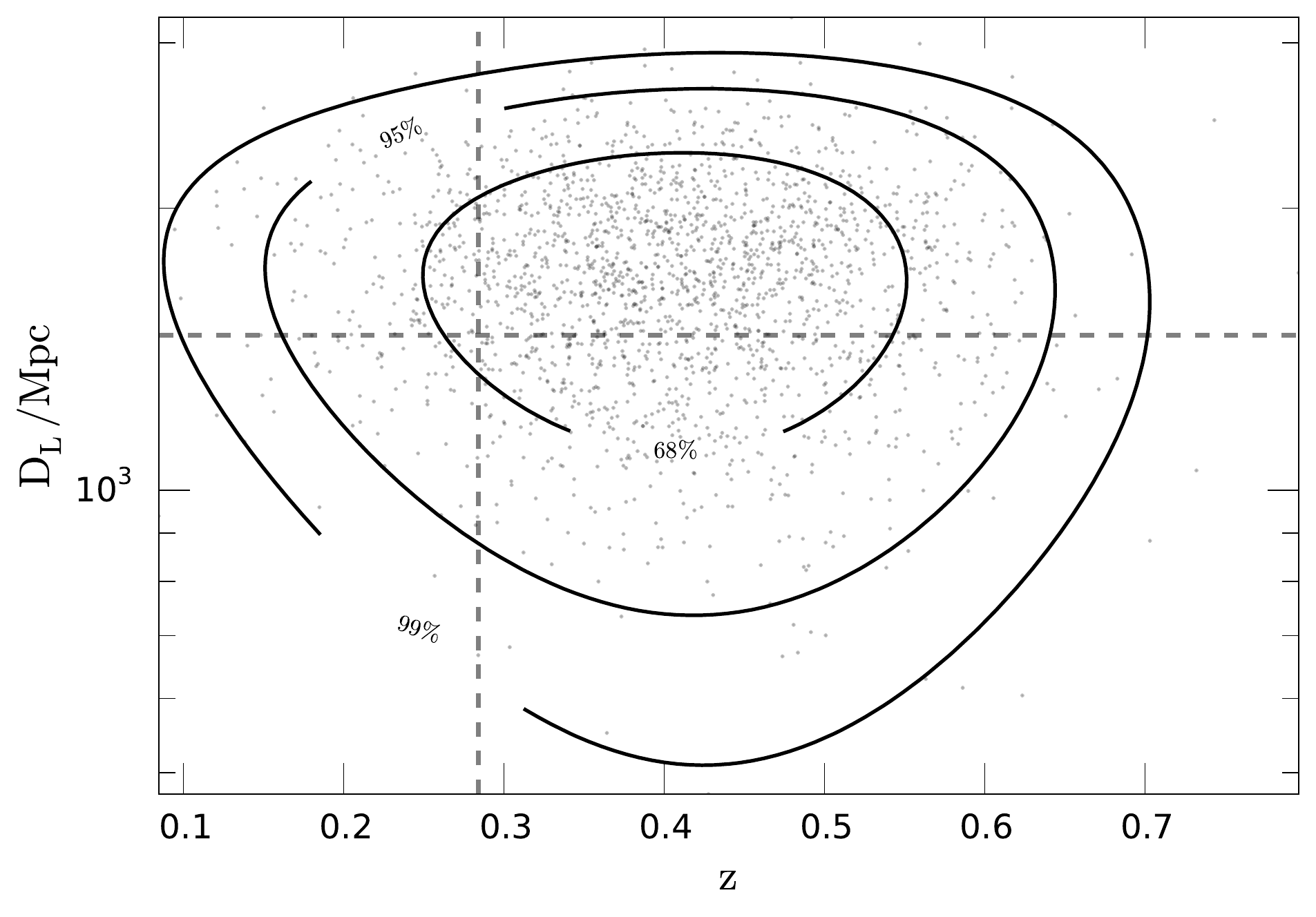}
	\includegraphics[width=\columnwidth]{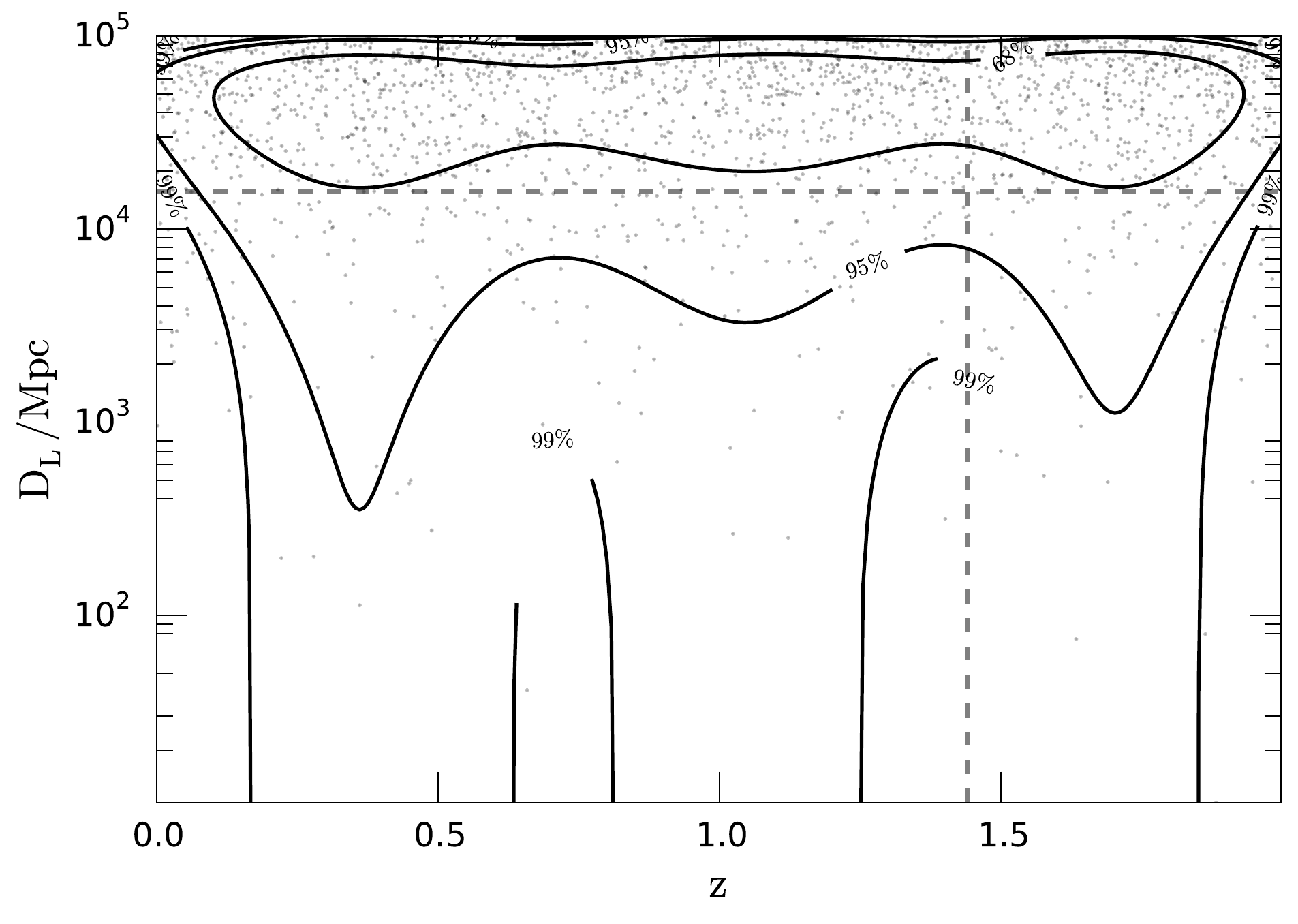} 
	\caption{Sample marginal likelihoods (\cref{eq:step1-likelihood}) for $z$
	and $D_L$. The left panel shows the marginal likelihood for an event with a
	network SNR=25. The right panel shows the marginal likelihood for an event
	with a network SNR=4. On both panels, the lines indicate the source distance
	and redshift.
}
\label{fig:marginal likelihoods}
\end{figure*}

A possible treatment of the problem would be to use the samples from
\cref{eq:step1-likelihood} and approximate it using a normalised histogram.
This procedure was successfully used in other unrelated studies
\cite{DelPozzo2011a}, however, for our purposes it is not accurate enough. In
fact, an histogram representation is dependent on a parameter, the bin size (or
equivalently the number of bins once the range is specified), which cannot be
inferred from the data but has to be chosen arbitrarily.
The majority of the quasi-likelihoods in \cref{eq:step2-likelihood}
tend to be very uniform over the cosmological parameter space for
individual sources and, as noted in Section \ref{s:AFM}, the inference
is strongly dominated by the prior on $z$.
Most sources are close or below the detection threshold of
the detector, thus a single source, in general, yields very little information about the
underlying cosmology, therefore any small fluctuation in the histogram
approximation due to the random variation of the number of samples in any
specific bin would be amplified and would eventually lead to a biased estimate
of the posterior probability density for $\cosmoP$. As an example,
compare the panels in Fig.~\ref{fig:marginal likelihoods}. The left
panel shows samples from Eq.~(\ref{eq:step2-likelihood}) for a source
having a network SNR=25. In this case, the isoprobability contours are
almost consistent with a normal distribution. The right panel shows
instead a source with network SNR=4. In this case the samples are
almost uniformly distributed, thus if one were to approximate
Eq.~(\ref{eq:step2-likelihood}) with a two dimensional histogram,
different choices of the bin size would result in different
approximations, which would yield to very different inference of $\cosmoP$.
Instead, we decided to follow a different course of action.  Given a set of
samples, for \emph{any} partition of the parameter space, the resulting
probability distribution of the observations is always a multinomial
distribution, therefore the ``probability distribution'' of the occurrences in
each bin is a Dirichlet distribution. The above property defines a Dirichlet
Process \cite{Ferguson1973a} which can be used to define an analytical
representation of the underlying probability distribution of which we have only
a finite number of samples available. For the mathematical details and
definitions, the reader is referred to the original paper by Ferguson
\cite{Ferguson1973a} or to the more recent discussions in \cite{Hjort2010a}.
We used the approximate variational algorithm \cite{Blei2006a} as implemented
in \cite{urlDPGMM} to find the Dirichlet Process Gaussian Mixture Model to
represent the integrand in \cref{eq:step2-likelihood}. The output of this
procedure is an analytical representation of the target probability
distribution as an infinite mixture of Gaussian distributions which is
analytical and continuous.  This form can then be used as the integrand in
\cref{eq:step2-likelihood} and the integral itself can be evaluated using
another Nested Sampling algorithm whose output can then be combined to compute
the posteriors for $\cosmoP$, \cref{eq:joint-posterior}.

For this last marginalisation, we assume priors on $\cosmoP$ that are
uniform in all parameters. In particular, $h \in [0.1,1.0]$,
$\Omega_m\in[0.0,1.0]$, $\Omega_\Lambda\in[0.0,1.0]$, $w_0\in[-2,0]$
and finally $w_1\in[-1,1]$. 

\section{Results}\label{s:results}

For each of the equations of state assumed we considered various scenarios 
within the set of $(h,\Omega_m,\Omega_\Lambda,w_0,w_1)$. We show
posterior distributions for the cosmological parameters obtained from
the joint analysis of $1,000$ events. We also report confidence
intervals from number of events greater than $1,000$ obtained via extrapolation.

Moreover, since all EOS yield very similar results,
we choose to report only posteriors for the cosmological parameters obtained
from the MS1 equation of state.

\begin{figure}[htp!]
	\includegraphics[width=\columnwidth]{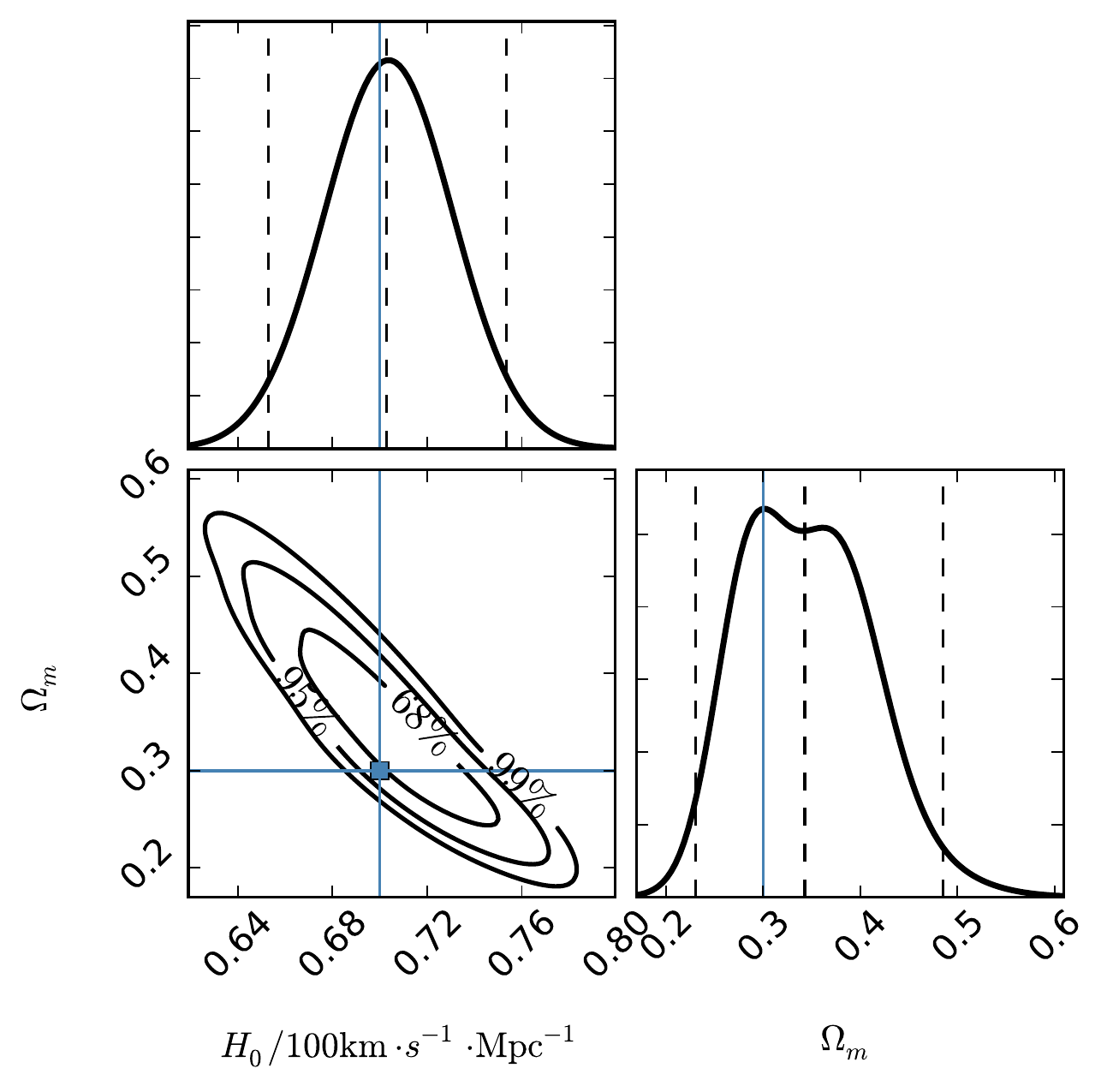}
	\caption{Posterior distributions for $h$ and $\Omega_m$ obtained from the
		analysis of 1,000 BNS events for a flat universe ($\Omega_k = 0$) and no dark
		energy equation of state. In the one dimensional
                posteriors the dashed lines indicate
                the 2.5\%, 50\% and 97.5\% confidence levels. In the two
                dimensional posterior distribution we show the 68\%,
                95\% and 99\% confidence regions. On all panels the solid (blue)
              lines indicate the fiducial values.}
	\label{fig:pos-2params}
\end{figure}

\begin{figure*}[htp!]
	\includegraphics[scale=0.9]{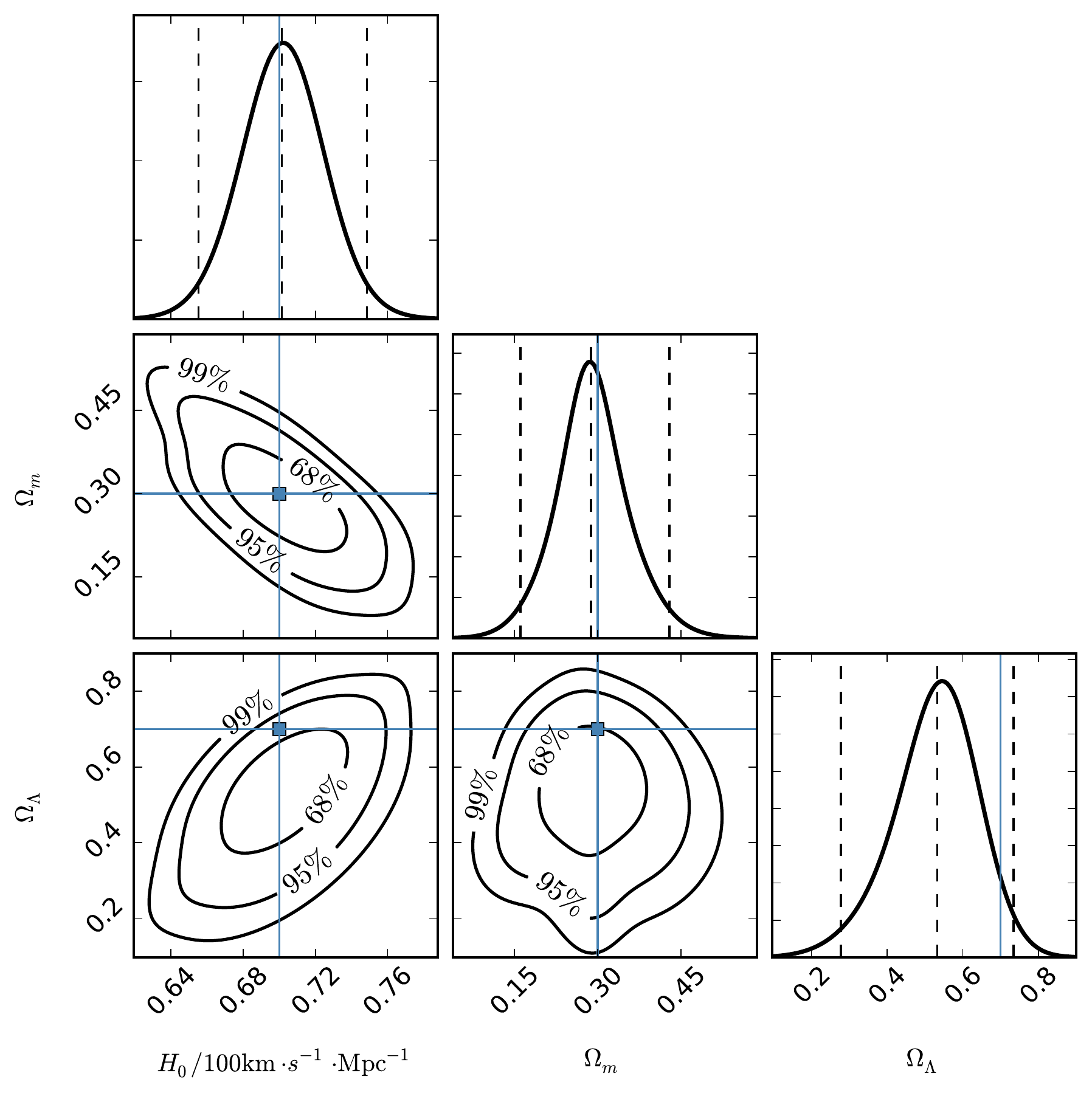}
	\caption{Same as Fig.~\ref{fig:pos-2params}, but for a general
        FRWL universe.}
	\label{fig:pos-3params}
\end{figure*}

\begin{figure*}[htp!]
	\includegraphics[scale=0.6]{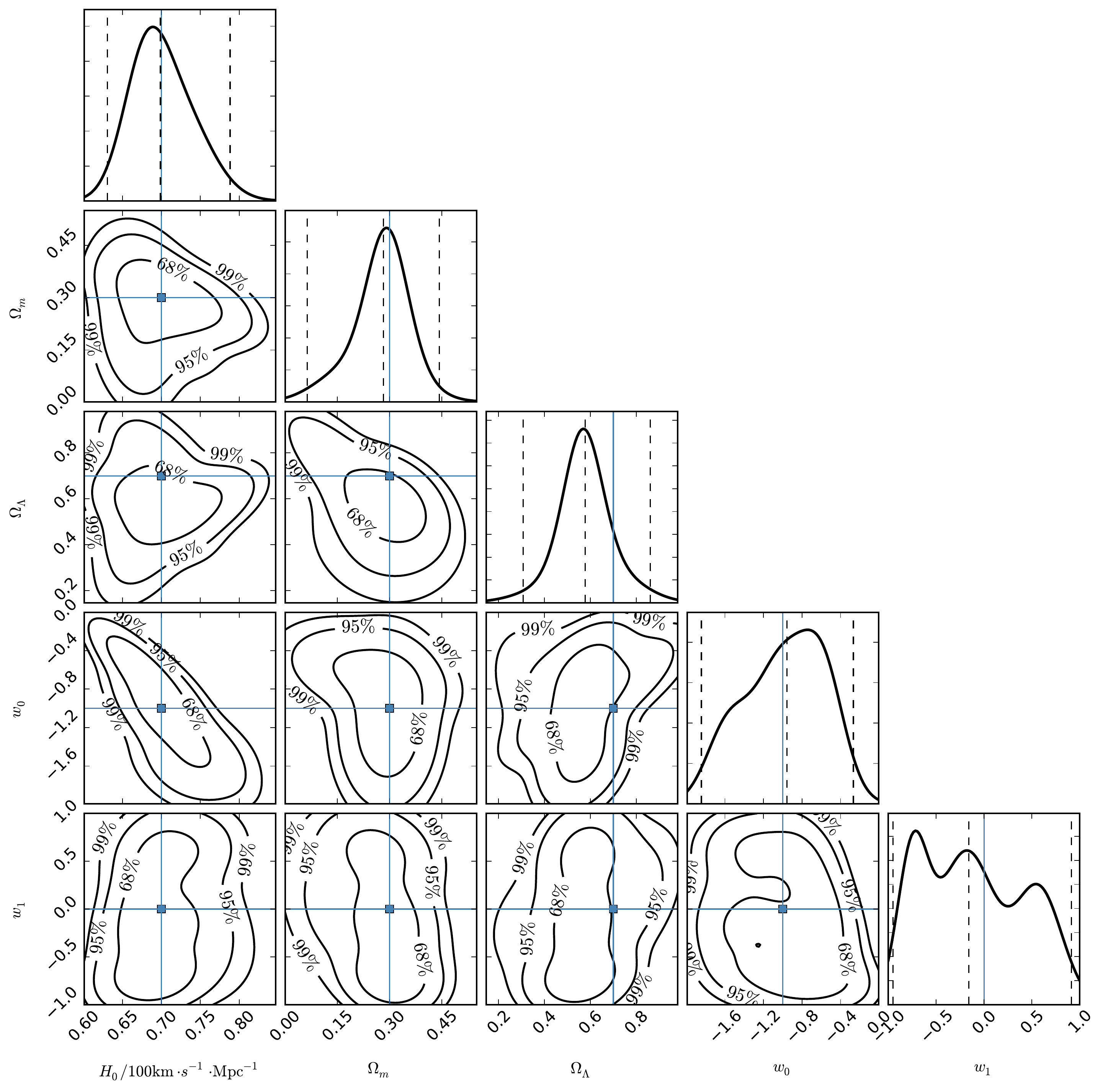}
	\caption{Same as Fig.~\ref{fig:pos-2params}, but for a general
        FRWL universe + DE parameters.}
	\label{fig:pos-5params}
\end{figure*}

We computed posterior distribution functions for three distinct cases:
\begin{enumerate}[i.]
\item flat FRWL universe, Fig.~\ref{fig:pos-2params};
\item general FRWL universe, Fig.~\ref{fig:pos-3params};
\item general FRWL universe, with Dark Energy parameters, Fig.~\ref{fig:pos-5params}.
\end{enumerate}

In the following subsections we report on the values of the various
cosmological parameters in each different cosmological model we
considered. We note here that, as expected, the accuracy of the
cosmological parameters measurement is better for the flat case and
get progressively worse with the increasing dimensionality of the
model under consideration. Also, all uncertainties we report are at
the $95\%$ confidence level.

Posterior distributions on the parameters of all cosmological models from the analysis of $10^3$ BNS
events are reported in Figs.~\ref{fig:pos-2params},
\ref{fig:pos-3params} and \ref{fig:pos-5params}. 

Depending on the actual astrophysical rate, ET will observe between
$10^3$ and $10^7$ BNS events per year~\cite{ETconcept}. It is computationally
unfeasible at the moment to simulate and analyse in a realistic way
such a large number of events. We therefore extrapolated the results
we obtained for 1,000 sources to the maximum expected number of events. 
We assumed that the central limit theorem holds, in other words that
our posteriors for $10^3$ events are approximately Gaussian, and
simply scaled the variance of the one dimensional posteriors by the
number of sources $N$. Tables~\ref{tab:summary-h},\ref{tab:summary-om} and \ref{tab:summary-w} show the extrapolation
of the 95\% widths (2$\sigma$) to a number of events
ranging from $10^4$ to $10^7$ for all relevant parameters for each of the cosmological models we
considered in our analysis.

\subsection{$H_0$}

We find that
$10^3$ BNS observations yield the following results:
for the model (i) the accuracy is $0.05(7\%)$  which remains approximately
constant in the general case (ii) and worsens to $0.08 (11\%)$ in the general
case (iii). In comparison with other GW studies, we find our
measurements to be significantly worse. For instance
\cite{DelPozzo2012a} reports a $95\%$ accuracy of $\sim 10\%$ with
second generation interferometers and similarly so do
\cite{Nissanke2013a} and \cite{Taylor2012b}. In comparison to
traditional methods, the most constraining measurement from the Planck
experiment in
conjuction with other methods reports an accuracy of $\sim 0.5\%$
\cite{Planck2015a}. So with $10^3$ GW sources current measurements are
far more accurate than those obtained with our method using ET. However, when
extrapolating to the potential number of observable BNS systems, we find
the Planck-like accuracy is reached with $\sim 10^5$ BNS observations. A
further order of magnitude improvement is seen for $10^7$ BNS
observations, see Table~\ref{tab:summary-h}.

\begin{table*}
	\centering
	\begin{tabular}{lccccc}
		\hline \hline
		\multicolumn{4}{c}{$\Delta h$}\\
		\diagbox{Model}{N} & $10^3$
                & $10^4$ & $10^5$ & $10^6$ & $10^7$\\
		\hline\\
		Flat FRW  & $0.5\times 10^{-1}$ & $1.6\times 10^{-2}$ & $0.5\times 10^{-2}$ & $1.6\times 10^{-3}$ & $0.5\times 10^{-3}$\\
		General FRW & $4.6\times 10^{-2}$ & $1.5\times 10^{-2}$ & $4.6\times 10^{-3}$ & $1.5\times 10^{-3}$ & $4.6\times 10^{-4}$\\
                General FRW+DE & $0.8\times 10^{-1}$ & $2.5\times 10^{-2}$ & $0.8\times 10^{-2}$ & $2.5\times 10^{-3}$ & $0.8\times 10^{-3}$\\
		\hline \hline
	\end{tabular}
	\caption{95\% accuracies on the measurement the reduced Hubble parameter
		for various detected numbers of sources for
                the general 5 parameter case. For $10^3$ sources,
                the widths have been computed using our Nested
                Sampling algorithm, otherwise the widths are the
                result of an extrapolation.}
	\label{tab:summary-h}
\end{table*}

\subsection{$\Omega_m$ and $\Omega_\Lambda$}

We find that with $10^3$ BNS events $\Omega_m$ can be measured with an
accuracy of $0.125(47.5\%),0.135(45\%)$ and $0.19(65.5\%)$ for the models (i), (ii)
and (iii) respectively. The above numbers compare very unfavorably
with the $\sim 2\%$ yield by Planck\cite{Planck2015b}. However, if
more sources are observed the current accuracy is reached with
$10^{6-7}$ sources, depending on the actual cosmological model. 

The situation is similar for the measurement of the cosmological
constant $\Omega_\Lambda$. With $10^3$ BNS observations we find an
accuracy of $0.23(37.5\%)$ and $0.275(39\%)$ for the models (ii)
and (iii) respectively. As a comparison Planck reports $\sim 0.9\%$ \cite{Planck2015a}.
However, a similar uncertanty is reached  with
$10^{6-7}$ sources. The results are summerised in Table~\ref{tab:summary-om}.

\begin{table*}
	\centering
	\begin{tabular}{lccccc}
		\hline \hline
		\multicolumn{4}{c}{$\Delta \Omega_m$}\\
		\diagbox{Model}{N} & $10^3$
                & $10^4$ & $10^5$ & $10^6$ & $10^7$\\
		Flat FRW & $1.3\times 10^{-1}$ & $4.0\times 10^{-2}$ & $1.3\times 10^{-2}$ & $4.0\times 10^{-3}$ & $1.3\times 10^{-3}$\\
		General FRW & $1.3\times 10^{-1}$ & $4.2\times 10^{-2}$ & $1.3\times 10^{-2}$ & $4.2\times 10^{-3}$ & $1.3\times 10^{-3}$\\
                General FRW+DE & $1.9\times 10^{-1}$ & $0.6\times 10^{-1}$ & $1.9\times 10^{-2}$ & $0.6\times 10^{-2}$ & $1.9\times 10^{-3}$\\
		\hline\\
		\multicolumn{4}{c}{$\Delta \Omega_\Lambda$}\\
		\diagbox{Model}{N} & $10^3$
                & $10^4$ & $10^5$ & $10^6$ & $10^7$\\
		\hline\\
		General FRW & $2.3\times 10^{-1}$ & $0.7\times 10^{-1}$ & $2.3\times 10^{-2}$ & $0.7\times 10^{-2}$ & $2.3\times 10^{-3}$\\
                General FRW+DE & $2.8\times 10^{-1}$ & $0.9\times 10^{-1}$ & $2.8\times 10^{-2}$ & $0.9\times 10^{-2}$ & $2.8\times 10^{-3}$\\
		\hline \hline
	\end{tabular}
	\caption{95\% accuracies on the measurement the matter energy
          density $\Omega_m$ and the cosmological constant $\Omega_\Lambda$
		for various detected numbers of sources for
                the general 5 parameter case. For $10^3$ sources,
                the widths have been computed using our Nested
                Sampling algorithm, otherwise the widths are the
                result of an extrapolation.}
	\label{tab:summary-om}
\end{table*}

\subsection{Dark Energy parameters}

In the case in which the $D_L - z$ relation is
modified to allow for a time varying cosmological constant, see
Eq.~(\ref{eq:DE}), the parameters $w_0$ and $w_1$ were
included in the model. From the analysis of $10^3$
events, at 95\% confidence, we find $\Delta w_0 = 0.8$ and $\Delta w_1 = 0.9$. In
comparison, assuming $10^3$ BNS events with optical counter parts and electromagnetic priors
on the remaining cosmological parameters\cite{Zhao2011a} finds
$\Delta w_0\approx 0.1$ and $\Delta w_1=0.3$.  Most recent
determination by Planck \cite{Planck2015b} of the parameters $w_0$ and
$w_1$ reports $\Delta w_0 \approx
0.2$ and $\Delta w_1 \approx
0.5$. Extrapolation to the potential number of sources observable in 1 year shows that the accuracy on $w_0$ and $w_1$
can be improved by 2 orders of magnitude, thus 1 order of magnitude
better than the current best estimates, see
Table~\ref{tab:summary-w}. It is worth noting that the posterior
distributions for $w_0$ and
$w_1$ are not very Gaussian, therefore the extrapolations to a large
number of events might not be as reliable as for the other
cosmological parameters.

\begin{table*}
	\centering
	\begin{tabular}{lccccc}
		\hline \hline
		\multicolumn{4}{c}{$\Delta w_0$}\\
		\diagbox{Model}{N} & $10^3$
                & $10^4$ & $10^5$ & $10^6$ & $10^7$\\
		\hline\\
                General FRW+DE  & $0.8\times 10^{0}$ & $2.5\times 10^{-1}$ & $0.8\times 10^{-1}$ & $2.5\times 10^{-2}$ & $0.8\times 10^{-2}$\\
		\hline \hline
		\multicolumn{4}{c}{$\Delta w_1$}\\
		\diagbox{Model}{N} & $10^3$
                & $10^4$ & $10^5$ & $10^6$ & $10^7$\\
		\hline\\
                General FRW+DE & $0.9\times 10^{0}$ & $2.9\times 10^{-1}$ & $0.9\times 10^{-1}$ & $2.9\times 10^{-2}$ & $0.9\times 10^{-2}$\\
		\hline \hline
	\end{tabular}
	\caption{95\% accuracies on the measurement the Dark Energy
          parameters $w_0$ and $w_1$
		for various detected numbers of sources. For $10^3$ sources,
                the widths have been computed using our Nested
                Sampling algorithm, otherwise the widths are the
                result of an extrapolation.}
	\label{tab:summary-w}
\end{table*}

\section{Discussion}\label{s:discussion}

In this study we investigated the potentialities of BNS observations with 
ET as cosmological probes. In particular, we quantified the cosmological information that can be 
extracted from pure GW observations of BNS. The ingredient that allows the measurement of the redshift 
is the knowledge of the NS equation of state and thus of the NS tidal deformability.

We simulated 1,000 events and relied on extrapolation to the expected $10^4 - 10^7$ events per
year. The main result of this study is that from GW alone, ET could measure all cosmological parameters with an
accuracy that is comparable with current state-of-the-art measurements
from EM missions.

This is the very first study of this kind, and therefore it should be regarded
as a proof-of-principle. Our analysis relies on a set of simplifying
assumptions which should be progressively relaxed for a comprehensive
investigation.  We neglected the NS spins and eventual precession of the
orbital plane. We do not expect this to be a serious limitation, since NS are
expected to be slow rotators \cite{OShaughnessy2008a}, however it is not clear
how a small degree of precession would impact the
analysis of ET data. There is evidence that for binary black holes and
NS black hole binaries accounting for precession of the orbital plane leads to more
accurate distance measurements \cite{VitaleEtal:2014L}. If a redshift
measurement can be associated to these classes of sources, the
determination of $\cosmoP$ would improve substantially, also thanks to
the considerably larger volume that is observable by ET.

Due to its low frequency sensitivity, BNS
signals in the local Universe  ($z<1$) will be in the ET sensitive band for a
time scale of hours to days, depending on the observer frame
masses. It is clear that this problem cannot currently be investigated
using a fully realistic simulation 
since the generation of the most accurate waveforms and cutting edge
data analysis algorithms are not yet sufficiently fast.  
A further complication
arises from the number of signals itself; a detection rate of $10^7$
events per year implies an average time delay between signals of $\sim$ 3
seconds. Given the duration of the signals in band, it follows that several
signals would be present simultaneously in the ET data stream at any given time. 
No systematic study nor algorithm yet exists to investigate this problem.
 
We further assumed perfect knowledge of the NS EOS. While there are
indications that second generation interferometers could measure the EOS
\cite{Read2009a,Hinderer2010a,Lackey2014a,DelPozzo2013a}, it is unlikely that
we would know it without any form of uncertainty. However, we do not consider
this a serious limitation as long as the error bar on any given value of the NS
mass $m$ and its tidal deformability $\lambda(m)$ is sufficiently small to
avoid confusion between different EOS. Moreover, we did not include in
our analysis the merger part of the waveform, which, especially for
the most distant sources, would be in the sensitive band of
ET. The inclusion of the merger in the analysis would yield
more information about the BNS mass and spin parameters as well as
introducing a possible further constraint on its redshift
\cite{MessengerEtAl:2013}, allowing for more precise measurements.  

A further element that deserves future investigation is the
effect of detection thresholds. We give a formal solution to the
problem of the inference of the cosmological parameters in Appendix
\ref{app:A}. However, we did not explore the details of its practical
implementation, which we defer to a future study.

We ignored the effects of the detector calibration uncertainties over the inference of the GW event parameters as well 
their impact over the global inference of $\cosmoP$. At the end of the first observing run of Advanced LIGO, 
typical amplitude uncertainties (which is relevant for the determination of $D_L$) and phase uncertainties (relevant for the estimation of $z$) 
were estimated at $\sim 10\%$ and $10$ degrees, respectively~\cite{GW150914-CALIBRATION}. 
Simulations indicate that ignoring the presence of such calibration errors does not lead to significant bias in the estimation of the GW parameters, as long as the SNR
is not very large~\cite{VitaleEtAl:2012}. However, data analysis models for GW analysis that marginalise over 
calibration uncertainties are now available~\cite{FarrEtAl:2015} and routinely utilised for the actual analysis~\cite{GW150914-PARAMESTIM}.
The additional calibration uncertainty increase the statistical uncertainty of the inferred parameters by a similar amount. 
In our case, the largest source of uncertainty would come from the amplitude calibration and thus on the determination of $D_L$ for the GW sources. 
Assuming, naively, an ET uncertainty budget of $\sim 10 \%$, we estimate a similar degradation of our inference over $\cosmoP$.

We also ignored the effects of weak lensing. Weak lensing is a zero mean process \cite{Bartelmann2001a}, thus, 
when averaging over thousands of sources, it will not induce an overall bias in the estimate of $\cosmoP$. 
A proper account of the lensing uncertainty, would lead to similar conclusions as the detector calibration uncertainty.

Even with the caveats discussed above, our study shows that even considering exclusively information coming from GW 
alone (with no input from any EM association) ET is capable to fully probe the evolution of the Universe and determine
the value of $\cosmoP$ with reasonable accuracy.  We emphasise that our results apply 
to a pure GW-based inference of $\cosmoP$. A more accurate determination of $\cosmoP$ from GW-EM joint detections 
may be possible, thus the results presented in this study should be regarded as a lower-limit 
to what the actual potentiality of ET is as a cosmological probe. Nevertheless, we showed that GW alone can be a feasible 
complementary and cross-validating route to probe the dynamics of the Universe.

\section*{Acknowledgments} 
This work benefitted from stimulating discussions and comments from
Bangalore Sathyaprakash, Alberto Vecchio, John Veitch, Ilya Mandel and
Christopher Berry. We thank the anonymous referees for their comments and suggestions. 
The work was funded in part by a Leverhulme 
Trust research project grant. WDP is funded by the program ``Rientro dei Cervelli Rita Levi Montalcini".

\appendix
\section{Inference of the cosmological parameters in presence of
  censored data}\label{app:A}

We want to infer the value of the cosmological parameters $\cosmoP\equiv
\left(H_0,\Omega_m,\Omega_\Lambda,\ldots\right)$ given a set of
gravitational waves observations. 
Consider a catalogue of gravitational wave events
$\mathcal{E}\equiv\{\event_1,\ldots,\event_N\}$. Each event is defined as a
stretch of data $d_i(t)$ given by the sum of noise $n_i(t)$ and a gravitational
wave signal $h_i(\vec{\Theta};t)$, where $\T$ indicates the set of all parameters of the signal and such
that the SNR $\rho$ is greater than a given
threshold $\rho_{\mathrm{th}}$.

The likelihood to observe the event $\event_i$ is given by $p(\event_i|\T,\mathrm{S},\rho_{\mathrm{th}},\info)$
where $\mathrm{S}$ is the signal model that relates the signal parameters
$\vec{\Theta}$ to a gravitational wave signal $h$.
The posterior distribution for the parameters in our signal model $\mathrm{S}$
comes from the application of Bayes' theorem
\begin{align} \label{eq:posterior}
	p(\T|\event_i,\mathrm{S},\rho_{\mathrm{th}},\info) \propto p(\T|\mathrm{S},\info) p(\epsilon_i | \T,\mathrm{S},\rho_{\mathrm{th}}, \info)
\end{align}
where $p(\T|\mathrm{S},\info)$ is the prior probability distribution for the
parameters $\T$. 
Given a certain cosmic coalescence rate $R(\cosmoP,z)$, there will be a certain
number $M\equiv M(\cosmoP,z)$ of gravitational wave events that will \emph{not} be registered as
events since their SNR will be below the selected 
threshold. Nevertheless, they encode information regarding the
Universe and they must be taken into account in our inference. 
The likelihood $\mathcal{L}^-_k $ for a non-detected event $\event^-_k$ is given by
\begin{align}
\mathcal{L}^-_k(\event^-_k,\rho_\mathrm{th})&\equiv
p(\event^-_k|\cosmoP,R(\cosmoP,z),\rho_\mathrm{th},\info) \\
&= \int_0^{\rho_{\mathrm{th}}}p(\event^-_k,\rho_k|\cosmoP,R(\cosmoP,z),\rho_\mathrm{th},\info)\di\rho_k\,.
\end{align}
For a set of $M$ non-detected events, the likelihood will be given by
\begin{align}
\mathcal{L}^-(\event^-,\rho_\mathrm{th})&=\prod_{k=1}^M
\mathcal{L}^-_k(\event^-_k,\rho_\mathrm{th})\\
&=\left[
  \mathcal{L}^-_k(\event^-_k,\rho_\mathrm{th})\right]^M
\end{align}

The number of non-detected events $M(\cosmoP,z)$ is a nuisance parameters which
depends on $\cosmoP$, the rate $R(\cosmoP,z)$, the detection threshold
$\rho_{\mathrm{th}}$, the observation time $T$ and the observed volume
$V(\cosmoP,\rho_{\mathrm{th}})$ as
\begin{align}
M(\cosmoP,z) = R(\cosmoP,z) V(\cosmoP,\rho_{\mathrm{th}}) T - N\,.
\end{align}
We are now in the position of writing the posterior distribution for
the cosmological parameters $\cosmoP$:
\begin{widetext}
\begin{align}\label{eq:full-like}
p(\cosmoP|\mathcal{E},N,\rho_{\mathrm{th}},S,\info)\propto
p(\cosmoP|S,\info) \int_0^{R_\mathrm{max}(\cosmoP,z)}\di R(\cosmoP,z)\, p(R(\cosmoP,z)|\cosmoP ,S,\info) \prod_{i=1}^N
\mathcal{L}(\event_i,\cosmoP)\sum_{M=0}^{\infty}\left[
  \mathcal{L}^-_k(\event^-_k,\rho_\mathrm{th})\right]^M  p(M|\cosmoP,R(\cosmoP,z),\rho_{\mathrm{th}})\, .
\end{align}
\end{widetext}

It is interesting to verify that Eq.~(\ref{eq:full-like}) reduces to
Eq.~(\ref{eq:joint-posterior}) in the limit of
$\rho_\mathrm{th}\rightarrow 0$. In this limit we have also
\begin{align}
&M\rightarrow 0 \\
&p(M|\cosmoP,R(\cosmoP,z),\rho_{\mathrm{th}})\rightarrow \delta_{M,0} \\
&\int_0^{\rho_{\mathrm{th}}}p(\event^-_k,\rho_k|\cosmoP,R(\cosmoP,z),\rho_\mathrm{th},\info)\di\rho_k
\rightarrow 0
\end{align}
therefore the term 
\begin{align}
\left[
  \mathcal{L}^-_k(\event^-_k,\rho_\mathrm{th})\right]^M \rightarrow 1
\end{align}
and the non-detection part of the likelihood reduces to 
\begin{align}
\sum_{M=0}^{\infty} \delta_{M,0} = 1\,.
\end{align}
Assuming that the rate $R(\cosmoP,z)$ is given by the integral of
Eq.~(\ref{eq:pz2}), we recover the form of the likelihood~(\ref{eq:joint-posterior})
which we used in our study.

%
%

\bibliography{CosmoWithNS_v2}
\end{document}